\documentstyle[graphicx]{aipproc}

\begin{document}

\title{New, Lead Free, Perovskites With a Diffuse Phase Transition:
NaNbO$_{3}$~ Solid Solutions}
\author{I. P. Raevski and S. A. Prosandeev}
\address{Department of Physics, Rostov State University, 5 Zorge Street,
344090, Rostov on Don, Russia}

\maketitle

\begin{abstract}
Some of ($1-x$)NaNbO$_{3}-(x)$ABO$_{3 }$~ perovskite solid
solutions exhibit a dramatic diffusion of the dielectric
permittivity $\varepsilon '$~ maximum and relaxor-type behavior
when the second component concentration exceeds a threshold value
$x_{0}$. The concentration phase transition to the relaxor-like
phase is abrupt (of the first order kind) that is seen from the
step in the dependence of the $\varepsilon'(T)$~ maximum
temperature, $T_{m}$, on $x$. The precursor of this transition is
a giant (up to 100 K) temperature hysteresis of $\varepsilon
'(T)$. Some relaxor-like properties appear even at $x < x_{0}$~ in
the course of cooling while disappear in the course of heating.
The experimental data obtained are qualitatively described within
a Landau-type phenomenological approach, assuming the relaxor-type
behavior to be local stress-induced.
\end{abstract}

\section{INTRODUCTION}
The relaxor behavior in perovskites was studied predominantly in
Pb-containing ternary compounds (PMN, PST, PSN) and solid
solutions (PLZT, PMN-PT) [1-3]. Last years some lead- free
BaTiO$_{3}$-derived solid solution compositions attracted much
attention as environmentally benign relaxor materials [4 ].
However up to now the obtained values of the dielectric
permittivity $\varepsilon'$~ maximum temperature, $T_{m}$, in the
BaTiO$_{3}$-based relaxors were too low for potential applications
[4]. The aim of the present paper is studying the dielectric
properties of some NaNbO$_{3}$ -based solid solutions forming a
new family of lead-free materials with diffuse phase transition
(DPT).

On heating NaNbO$_{3}$~ exhibits a series of six phase transitions
from the low temperature ferroelectric (FE) phase N ($R3c$) to the
high temperature paraelectric (PE) cubic phase $U$ ($Pm3m$)
through different antiferroelectric (AFE) and PE phases [5]. An
$\varepsilon(T)$ maximum originating from the first order
transition between two AFE phases: $P(Pbma)$ and $R(Pmnm)$
 is observed at 350-370$^{0}$C.

Similar to solid solutions of other perovskite antiferroelectrics,
the NaNbO$_{3}$ - ABO$_{3 }$~ binary solid solution systems can be
divided into two groups [6]. In the solid solutions of group I
(e.g. (Na,Li)NbO$_{3}$~ and (Na,K)NbO$_{3}$~ systems) the high
temperature FE phase appears at small (a few mol.$\%$) content
$x$~ of the second component ABO$_{3}$, the $T_{m}(x)$~ dependence
is rather smooth and the $\varepsilon'(T)$~ maxima are sharp. In
the solid solutions of group II the AFE phase remains stable up to
a comparatively high $x$~ values. In contrast to the solid
solutions of other perovskite antiferroelectrics, the $T_{m}(x)$~
dependence of the NaNbO$_{3}$-based group II solid solutions
remains smooth only up to a threshold $x=x_{0}$~ value. At $x >
x_{0}$~ the phase, usually referred to as FE (though sometimes it
is supposed to be  ferrielectric), becomes stable, which is
accompanied by an abrupt drop in the $T_{m}$~ values and dramatic
diffusion of the $\varepsilon '(T)$~ maximum. While the $T_{m}$~
values of the compositions with $x < x_{0 }$~ do not depend on
frequency, the compositions with $x > x_{0}$~ were reported to
exhibit a frequency dispersion of $\varepsilon '$~ and an increase
of $T_{m}$ with frequency [6]. Thus, it seems that NaNbO$_{3}$
-based solid solutions belonging to group II exhibit a
relaxor-like behavior at $x > x_{0}$. Besides, the lack of
systematic data on the properties of such materials prevents one
from definite conclusions. Below we will consider some group II
NaNbO$_{3}$- based solid solutions dielectric properties
dependence on the concentration of the second and third components
as well as on temperature and frequency.

\section{EXPERIMENTAL RESULTS}

NaNbO$_{3}$-Gd$_{1 / 3 }$NbO$_{3}$~ solid solution crystals were
grown by the flux method. The details of crystal preparation and
characterization have been described elsewhere [7]. The ceramic
samples of NaNbO$_{3}$ -based solid solutions were prepared by
routine solid-state reaction route. The density of the obtained
ceramics was about 90-95 $\%$~ of theoretical one. For dielectric
measurements Aquadag electrodes were deposited on the opposite
faces of the as-grown crystals, while Ag paste was fired to the
grinded disk surfaces of ceramic samples . The dielectric studies
were carried out in the 1 kHz -1 MHz range in the course of both
heating and cooling at a rate of 2 -3 $^{0}$ C/min with the aid of
the $R5083$ and $E7-12$ capacitance bridges.

Fig. 1 shows the typical evolution of the $\varepsilon '(T)$~
dependencies with the composition for the solid solution of group
II. The similarity of the data obtained for ceramic samples and
crystals shows that the peculiar properties of the group II solid
solutions are not due to the immiscibility effect typical of many
solid solution ceramics. A strong frequency dispersion of both
real ($\varepsilon '$) and imaginary ($\varepsilon ''$) parts of
complex dielectric permittivity observed at temperatures exceeding
$T_{m}$~ or $T''_{m}$~ (Fig. 2) is likely to be caused by a
decrease of the conductance influence with increasing frequency.
Though the $\varepsilon '(T)$~ maxima of the group II
NaNbO$_{3}$-based solid solution compositions with $x > x_{0}$~
are smeared and a frequency dispersion of $\varepsilon '$~ is
observed, the shift of $T_{m }$~ with frequency is usually much
smaller than in the case of PbMg$_{1 / 3}$Nb$_{2 / 3}$O$_{3
}$-type relaxors (Fig.2). Similar to other FE and AFE with DPT
[1-3], the permittivity of NaNbO$_{3}$- based solid solution
compositions with $x > x_{0 }$~ does not follow the Curie-Weiss
law in a broad temperature range above $T_{m}$~ (Fig. 2) while at
higher temperatures the Curie-Weiss behavior is observed. It is
well documented that in FE with DPT the extrapolated Curie-Weiss
temperature of the FE phase transition, $T_{CW}$, is necessarily
higher than $T_{m}$~ due to a large contribution to the
high-temperature permittivity from the regions that have a higher
transition temperature [1-3]. In contrast to this, in all the
group II NaNbO$_{3}$-based solid solutions studied, the $T_{CW }$~
values are much lower than $T_{m}$~ (Fig.2). Figure 3 shows the
concentration dependencies of $T_{m}$~ measured on heating for
some solid solutions belonging to group II. It is interesting to
note that for many solid solution systems of group II the linear
extrapolations of the $T_{m }- x$~ diagram portions from the $x >
x_{0}$~ region intersect at the same point ($x = 0$, $T \sim
150^{o}$C) corresponding to the minor $\varepsilon '(T)$ anomaly
often observed in NaNbO$_{3}$~ [8].

\begin{figure}[htbp]

\centerline{\includegraphics[width=4.34in,height=2.61in]{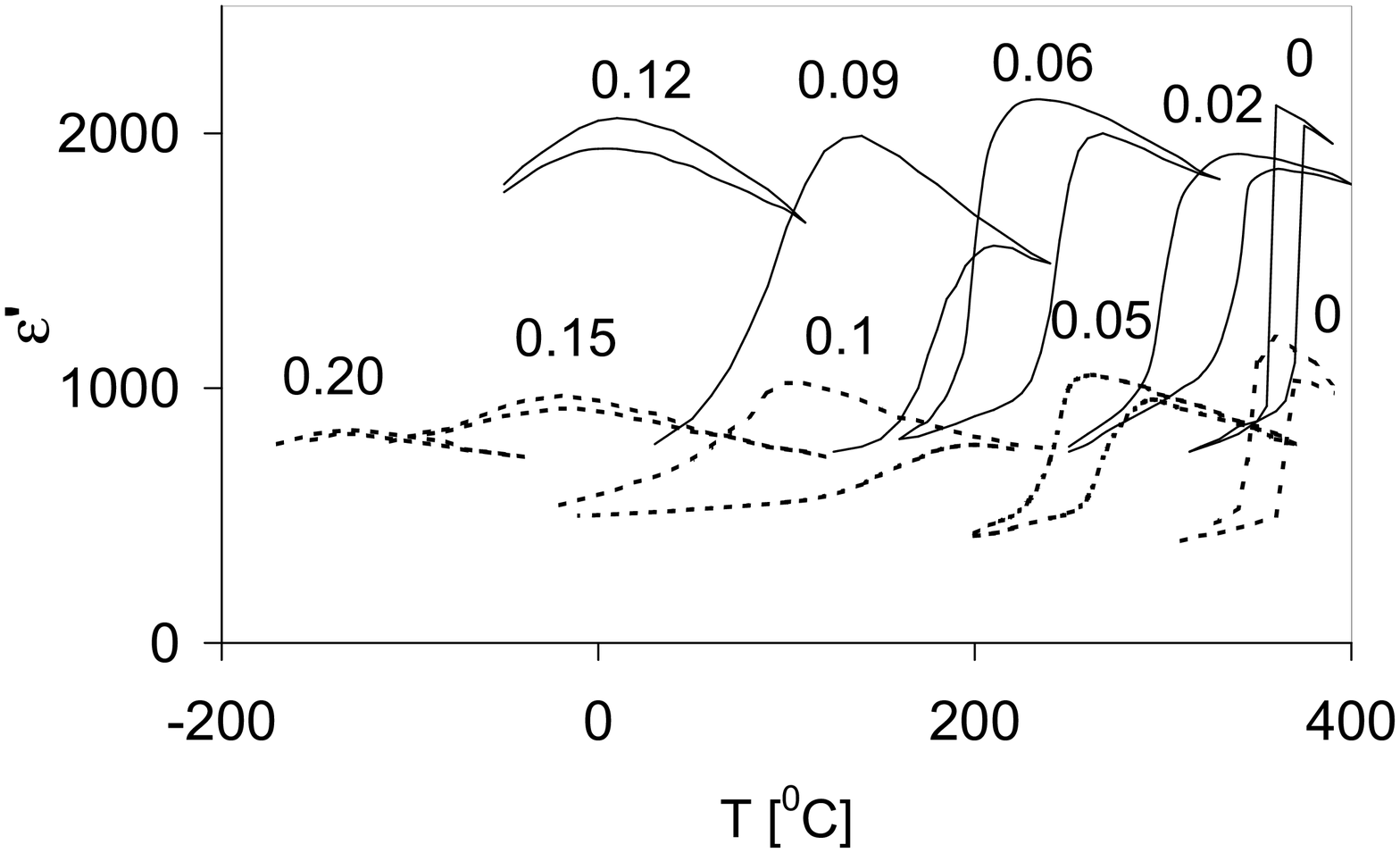}}
\caption{Evolution of the $\varepsilon $'(T) dependencies measured
on heating and subsequent cooling with composition for (1-x)
NaNbO$_{3}$-xGd$_{1 / 3 }$NbO$_{3}$~ single crystals (solid lines)
and ceramics (dashed lines). Figures correspond to $x$~ values.}
\label{fig1}
\end{figure}

\begin{figure}[htbp]
\centerline{\includegraphics[width=4.94in,height=2.69in]{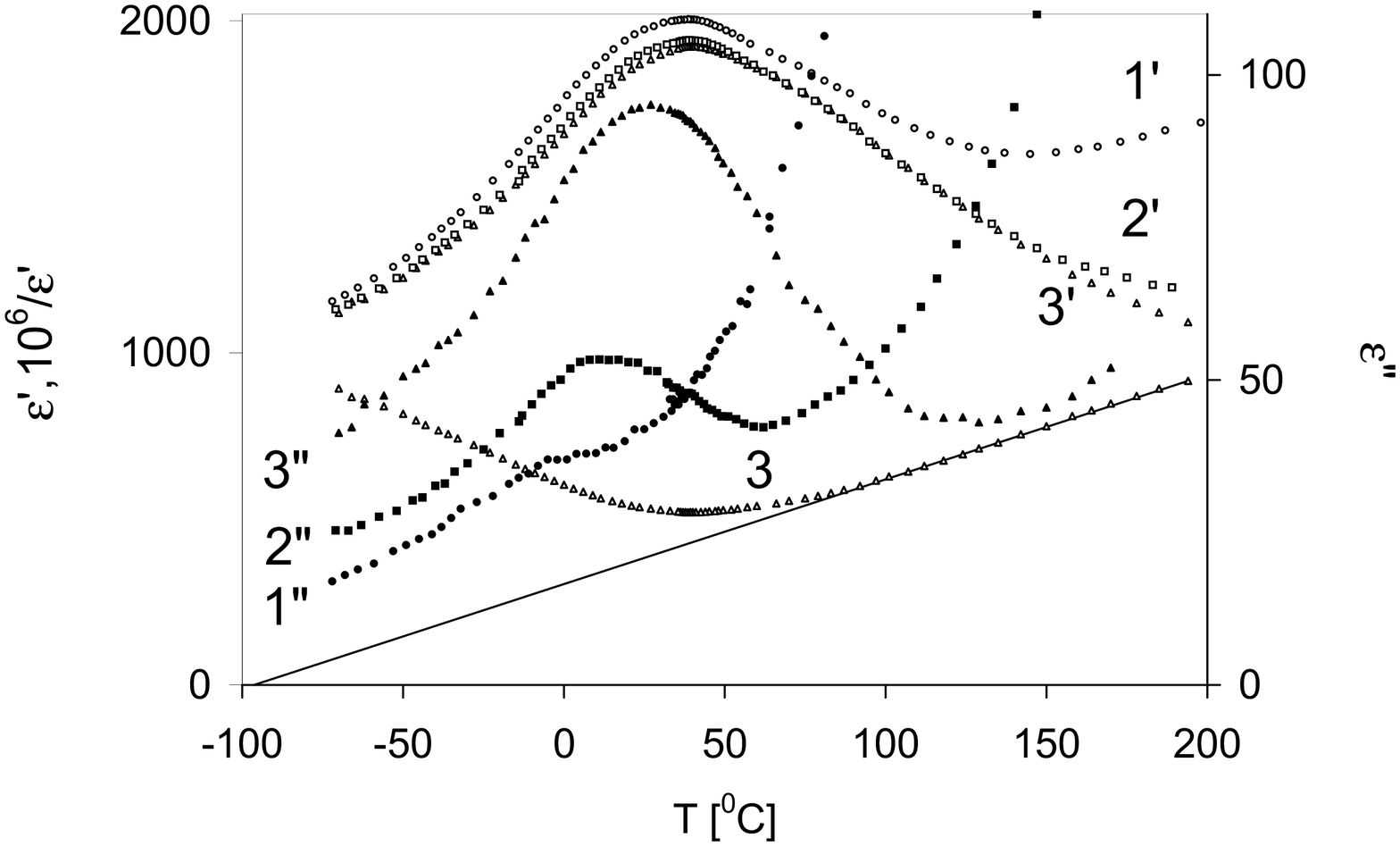}}
\caption{Temperature dependencies of $\varepsilon '$(1'-3'),
$\varepsilon ''$ (1''-3'') and 10$^{6}$/$\varepsilon '$~ (3)
measured at 1 kHz (1',1''), 10 kHz (2',2'') and 100 kHz (3,3',3'')
for 0.88NaNbO$_{3}$-0.12Gd$_{1 / 3 }$NbO$_{3}$~ single crystal}
\label{fig2}
\end{figure}

\begin{figure}[htbp]

\centerline{\includegraphics[width=4.49in,height=2.83in]{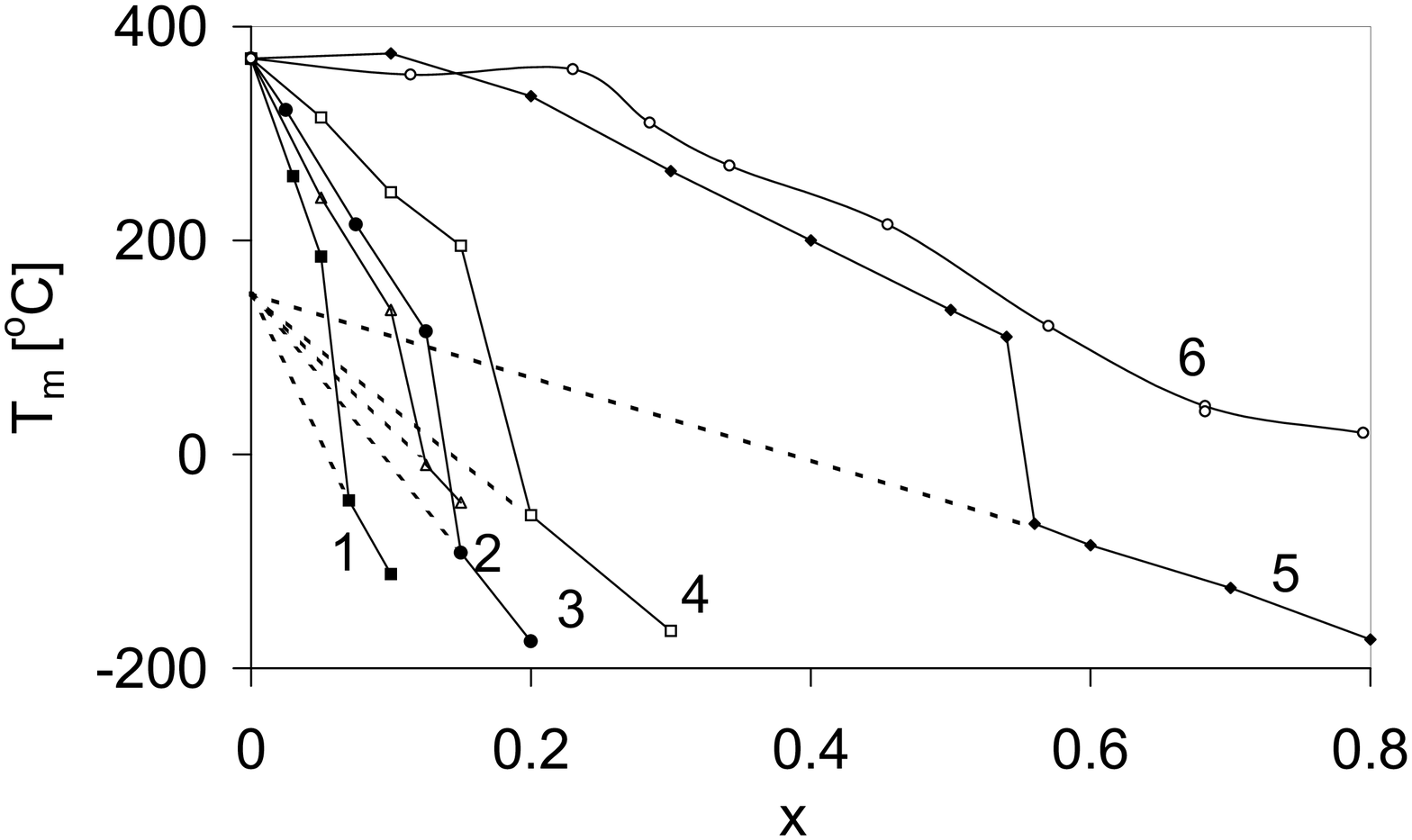}}
\caption{Concentration dependencies of the $\varepsilon $'(T)
maximum temperature T$_{m}$~ measured on heating for some
NaNbO$_{3}$ -based solid solutions belonging to the group II (1-5)
as well as for 0.88 NaNb$_{1 - x}$Ta$_{x}$O$_{3}$-0.12LiNbO$_{3}$~
[16] system (6). The second components of the solid solutions are:
1.BiFeO$_{3}$; 2. SrCu$_{1 / 3}$Nb$_{2 / 3}$O$_{3}$; 3.CaTiO$_{3
}$~ [17]; 4. SrTiO$_{3 }$~ [23]; 5. NaTaO$_{3 }$[24];}
\label{fig3}
\end{figure}

A specific feature of the group II solid solutions is an
anomalously large value of the thermal hysteresis \textit{$\Delta
$T}$_{h}$~ of the $\varepsilon '(T)$ dependence typical of the
compositions with $x < x_{0}$. The $\Delta T_{h}$~ values increase
with $x$~ and in some systems exceed 100 K for compositions
adjacent to $x_{0}$~ (Figs. 1, 4). The concentration dependence of
$\Delta T_{h}$~ is substantially nonlinear (Fig. 4). When a small
amount of FE third component is added to the given solid solution,
$\Delta T_{h}$~ values usually increase [10-12]. At higher FE
component content the $\Delta T_{h}$~ values decrease dramatically
[11]. If the temperature is stabilized in the course of cooling
within the thermal hysteresis range the temporal changes in
$\varepsilon '$~ do not exceed a few {\%} during several hours
[12]. Application of DC bias on compositions exhibiting giant
$\Delta T_{h}$~ values does not change the $\varepsilon '(T)$~
curve measured in the heating mode while in the course of
subsequent cooling some lowering of the $\varepsilon '(T)$~
maximim was observed [12,13]. Anomalously large values of $\Delta
T_{h}$~ were observed both in the ceramics and single crystals and
can serve as an experimental evidence of the fact that the given
solid solution belongs to group II [7,9,11]. It is worth noting
that in perfect NaNbO$_{3}$~ single crystals the $\Delta T_{h}$~
value does not exceed 10 K , but increases substantially with the
increase of the oxygen vacancy concentration [14]. In nominally
stoichiometric NaNbO$_{3}$~ ceramics the $\Delta T_{h}$~ values of
30 - 40 K are usually observed (Fig. 1). These larger values of
$\Delta T_{h}$~ in ceramic samples are likely to be attributed to
a high point-defect concentration, e.g. due to evaporation of
Na$_{2}$O during sintering. Indeed, in Na$_{2}$O-deficient Na$_{1
- x}$NbO$_{3 - x / 2 }$~ ceramics [15] the $\Delta T_{h}$~ values
are substantially larger than in the nominally stoichiometric
ones. Both the oxygen- and Na$_{2}$O -deficiency leads to an
increase of the $T_{m}$~ values in NaNbO$_{3}$~ [14,15] and an
almost linear correlation of $\Delta T_{h}$~ and $T_{m}$~ is
observed.

\begin{figure}[htbp]

\centerline{\includegraphics[width=4.36in,height=2.35in]{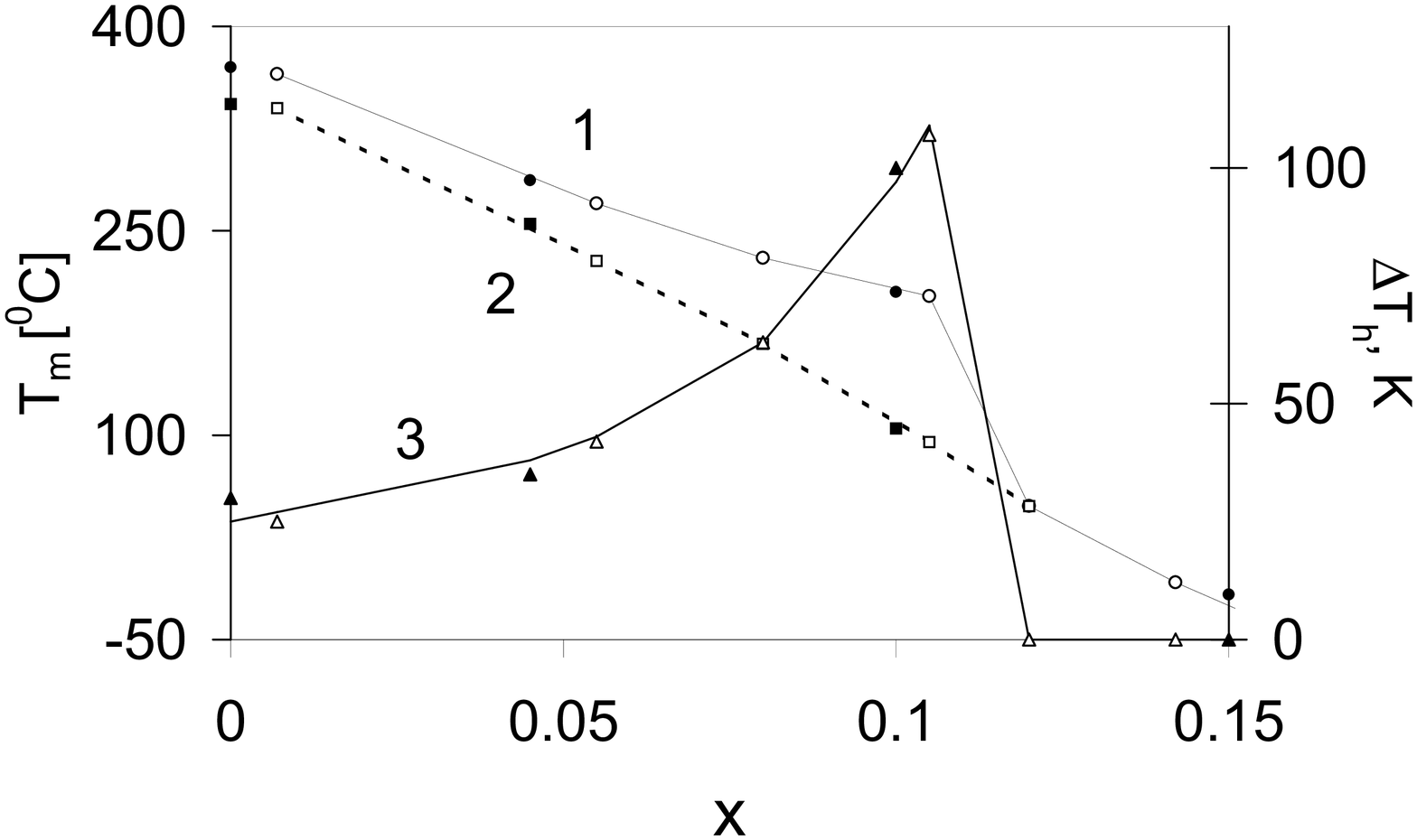}}
\caption{Concentration dependence of the $\varepsilon '(T)$~
maximum temperature $T_{m}$~ measured on heating (1) and on
cooling (2) as well as the $\varepsilon '(T)$~ temperature
hysteresis $\Delta T_{h}$~ values (3) for (1-x)NaNbO$_{3}$-xGd$_{1
/ 3 }$NbO$_{3}$~ crystals (open symbols) and ceramics (filled
symbols)} \label{fig4}
\end{figure}

\begin{figure}[htbp]

\centerline{\includegraphics[width=4.67in,height=2.64in]{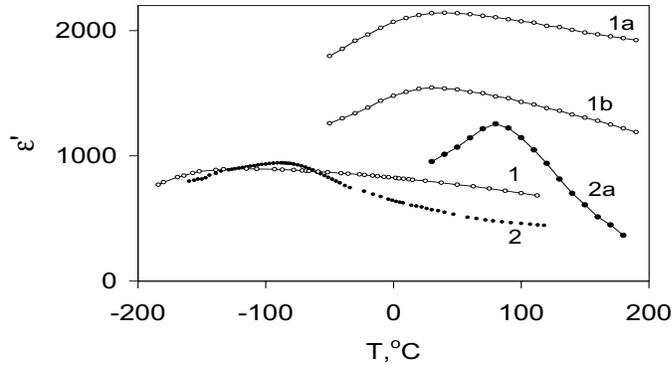}}
\caption{Changes in the $\varepsilon $'(T) dependencies of
0.8NaNbO$_{3}$-0.2Na$_{0.5}$Bi$_{0.5}$TiO$_{3}$~ (1) and
NaNb$_{0.45}$Ta$_{0.55}$O$_{3}$~ (2) solid solution ceramics
caused by addition of 10 mol{\%} of LiNbO$_{3}$~ (curves 1a and
2a) or KNbO$_{3}$~ (1b). Curve 2a is drawn using the data of Ref.
[16].} \label{fig5}
\end{figure}

It should be mentioned that a noticeable step in the $T_{m}(x)$~
dependence of the NaNbO$_{3}$ -based solid solutions belonging to
group II is observed only for the $T_{m}$~ values measured in the
heating mode. If $T_{m}$~ is determined from the $\varepsilon
'(T)$~ dependence measured in the course of cooling, the
$T_{m}(x)$~ curve is rather smooth (Fig. 4). The $\varepsilon
'(T)$~ dependencies measured in the cooling mode are usually more
diffused as compared to those measured upon heating (Fig.1).
Moreover, just at the step in the $T_{m}(x)$~ dependence the
$\varepsilon '(T)$~ curve obtained on cooling down has a
relaxor-like diffuse maximum that can imply that the relaxor-type
state had been already appeared in a metastable thermodynamic
state occupied at cooling. The stability of this state arises at
temperatures lower than the step in the $T_{m}(x)$~ dependence. In
the latter case (i.e. for the compositions with $x > x_{0})$~ the
curves obtained in the cooling and heating modes practically
coincide.

As one can see from Figs. 1 and 3, the $T_{m}$~ values of the
known binary NaNbO$_{3}$ -based relaxor-type solid solutions,
belonging to group II, are usually well below the room
temperature, and the maximal $\varepsilon '$~ values, $\varepsilon
'_{m}$, are much lower than in the Pb-containing relaxors.
However, in the ternary NaNbO$_{3}$-NaTaO$_{3}$-LiNbO$_{3}$~ solid
solution system a dramatic increase in both the $T_{m}$~ and
$\varepsilon '_{m}$~ values with the LiNbO$_{3}$~ content is
observed [16] for the compositions from the $x > x_{0}$~ range
(Figs. 3 and 5). Similar behavior is typical of other
NaNbO$_{3}$-based solid solutions (Fig. 5).

\section{DISCUSSION}
When discussing the experimental data we will use a
phenomenological Landau-type approach. For the sake of simplicity
we do not follow the complex symmetry of the NaNbO$_{3}$-based
solid solutions but rather, in order to make a qualitative
description, we consider a simplified case when there are two
nonpolar (AFE) ($Q_{1}$, and $Q_{2}$) and one polar (FE) ($P)$~
order parameters. An important constraint on the Free energy
expansion is that an AFE phase transition may exhibit a peak or
step in the $\varepsilon '(T)$ only because of coupling between FE
and AFE order parameters [18,19]. The impurities we consider are
expansive that implies that they produce local random stress
($\sigma _l$ ) leading to a local expansion of the lattice
parameter and, hence, to local strain ($e)$. The small observed
dispersion of $\varepsilon '(T)$~ at $x>x_{0 }$~ requires the
introduction of an additional parameter ($d$), which is the local
mean square impurity- induced dipole (or multipole) moment. We
assume that local expansion around a defect may lead to
noncentrosymmetric relaxations of neighboring atoms yielding a net
local moment, or a multipolar set of local moments. Averaged over
the whole crystal, these relaxations do not generate a net
macroscopic moment but local moments cause dispersion of
$\varepsilon '(T)$. Extended dipole-dipole interactions are
assumed to be sufficiently weak that they can be ignored.

We expand the Free energy as a sum of $x$-independent ($F_{1}$)
and $x$-dependent ($F_{2}$) contributions, $F = F_1 + xF_2$~ where

\begin{equation}
\label{eq2}
\begin{array}{l}
 F_1 = F_0 + \textstyle{1 \over 2}\alpha \,_1 (T)P^2 + \textstyle{1 \over
4}\beta _1 P^4 - EP + \\
 + \textstyle{1 \over 2}\alpha \,_{21} (T)Q_1^2 + \textstyle{1 \over
2}\alpha \,_{22} (T)Q_2^2 - \textstyle{1 \over 4}\beta _{\,21} Q_1^4 +
\textstyle{1 \over 4}\beta _{\,22} Q_2^4 + \textstyle{1 \over 6}\gamma
_{\,21} Q_1^6 + \\
 + \textstyle{1 \over 2}\beta _{\,121} P^2Q_1^2 + \textstyle{1 \over 2}\beta
_{\,122} P^2Q_2^2 \\
\\
F_2 = \textstyle{1 \over 2}c_L e^2 - \sigma _l e - \textstyle{1
\over 2}a_1 eP^2 + \textstyle{1 \over 2}a_{21} eQ_1^2 +
\textstyle{1 \over 2}a_{22} eQ_2^2 - \lambda de + \textstyle{1
\over 2}\kappa d^2 \end{array}
\end{equation}

\noindent Here $\alpha _i = c_1 (T - T_1^{(0)} )$, $\alpha _{2i} =
c_{2i} (T - T_{2i}^{(0)} )$, $T_1^{(0)} $~ and $T_{2i}^{(0)} $~
are the bare Curie temperatures for the FE and for AFE phase
transitions respectively, $E$~ is external electric field, $e$~ is
the symmetrical part of the strain tensor. From the equilibrium
condition with respect to $e$~ and $d$~ one has

\begin{equation}
\label{eq4}
\begin{array}{l}
e = \frac{1}{c_L }\left( {\sigma _l + \lambda d + \textstyle{1 \over 2}a_1
P^2 - \textstyle{1 \over 2}a_{21} Q_1^2 - \textstyle{1 \over 2}a_{22} Q_2^2
} \right)
\\
d = \lambda e/\kappa
\end{array}
\end{equation}

\noindent It implies that $d$~ multiplied by $\lambda $~ plays the
role of the local stress and hence it effectively enlarges the
local stress produced by impurities, $d$~ proves to be
proportional to the strain $e$. From (\ref{eq2}) and (\ref{eq4})
one can easily find that $e$~ (as well as $d$, see [20]) is
enhanced due to mutual coupling between $e$~ and $d$:

\begin{equation}
\label{eq6}
e = \frac{1}{c_L - \lambda ^2 / \kappa }\left( {\sigma _l + \textstyle{1
\over 2}a_1 P^2 - \textstyle{1 \over 2}a_{21} Q_1^2 - \textstyle{1 \over
2}a_{22} Q_2^2 } \right)
\end{equation}

\noindent By using (\ref{eq6}) one can rewrite Free energy in the
simple form containing expansions only with respect to $P$~ and
$Q$:

\begin{equation}
\begin{array}{l}
 F = F_0 ' + \textstyle{1 \over 2}A_1 \left( {x,T} \right)P^2 + \textstyle{1
\over 4}B_1 \left( x \right)P^4 - EP + \textstyle{1 \over 2}A_{21} \left(
{x,T} \right)Q_1^2 +
\\
\textstyle{1 \over 2}A_{22} \left( {x,T} \right)Q_2^2
 - \textstyle{1 \over 4}B_{21} \left( x \right)Q_1^4 + \textstyle{1 \over
6}\Gamma _2 \left( x \right)Q_1^6 + \textstyle{1 \over 4}B_{22}
\left( x \right)Q_2^4 +
\\
 \textstyle{1 \over 2}B_{121} \left( {x,T} \right)P^2Q_1^2 +
\textstyle{1 \over 2}B_{122} \left( {x,T} \right)P^2Q_2^2 + ... \\
 \end{array} \label{Free}
\end{equation}

\noindent where $F_0 ' = F_0 - x\sigma _l^2 / 2\varsigma $~ and
coefficients $A$~ and $B$~ depend on concentration, e.g.

\begin{equation}
\begin{array}{l}
B_{21} \left( x \right) = \beta _{21} + xa_{21}^2 / 2\varsigma
\\
B_{121} \left( x \right) = \beta _{121} + xa_1 a_{21} / 2\varsigma
\\
B_{122} \left( x \right) = \beta _{122} + xa_1 a_{22} / 2\varsigma
 \end{array} \label{eq8}
\end{equation}

\noindent Here $\varsigma = \left( {c_L - \lambda ^2 / \kappa }
\right)^2 / c_L $. We have obtained that the coupling of the order
parameter $Q_{i}$~ with the strain $e$~ produced by the impurities
leads to the dependence of the coefficients in the Landau
expansion on the concentration $x$~ (see also [21]).

Composition dependence in coefficients (\ref{eq8}) implies that
the Curie temperature also depends on $x$:

\begin{equation}
\begin{array}{l}
T_1 = T_1^{(0)} + x\sigma _l a_1 / c_1 \varsigma
\\
T_{2i} = T_{2i}^{(0)} - x\sigma _l a_{2i} / c_{2i} \varsigma
\end{array}
\label{eq15}
\end{equation}

\noindent From these expressions it is seen that if $a_1 > 0$~
and $a_2
> 0$~ then $T_{1}$~ increases with $x$~ and $T_{2}$~ decreases with $x$~
linearly. The latter result corresponds to the experimental data:
all group II NaNbO$_{3}$-based solid solutions studied exhibit a
decrease in $T_{m}$~ with $x$. The assumed signs of $a_{1}$~ and
$a_{2}$~ are natural as the electrostrictive constant $a_{1}$~ is
always positive (if in the F one puts the sign "-" in front of it)
and $a_2$~ should also be positive (if one puts the sign "+" in
front of it) because compression of the lattice ($e < 0$)
typically promotes AFE order while the expansion ($e > 0$)
promotes FE order.

From expansion (\ref{Free}) one can easily obtain: the critical
temperature of the AFE phase transition, $T_{Ai}$; the temperature
hysteresis width, $\Delta T_h $; and the jump of the AFE order
parameter, $\Delta Q_1 $:

\begin{equation}
\label{eq16}
\begin{array}{l}
 T_{A1} = T_{21} + 3B_{21}^2/16c_{21} \gamma_{21}
   - B_{121}P^2 /c_{21}
\\
 T_{A2} = T_{22} -
B_{122} P^2/c_{22}
\\
\Delta T_h = B_{21}^2 /4c_{21} \gamma _{21}
\\
\Delta Q_1 = \sqrt {B_{21} /2\gamma _{21}}
\end{array}
\end{equation}

\noindent From (\ref{eq15}) it is clear that T$_{2i}$~ is a linear
function of $x$, which implies that $T_{Ai}(x)$~ is also linear at
small $x$. This corresponds to the experimentally observed linear
decrease of $T_{m}(x)$~ with $x$~ both above and below $x_{0}$~ as
well as explains why linear extrapolations to the point $x=0$~ for
various second components, intersect at one point (Fig. 3). These
points correspond to the bare Curie temperatures $T_{2i}^{(0)}$.
The slopes of the linear dependencies above and below $x_{0 }$~
are different as are the electrostriction constants ($a_{21} >
a_{22} $). Thus: Curie lines $T_{A1}(x)$~ and $T_{A2}(x)$~
intersect at $x=x_{0}$; $Q_{1}$~ is nonzero for $T < T_{A1}$~ ($x
< x_{0}$); $Q_{2}$~ is nonzero for $T < T_{A2}$~ ($x_{0} < x$ ).
The point $x = x_0 $~ can be found from the equality: $T_{A1} (x)
= T_{A2} (x)$.

At the intersection of curves $T_{A1}(x)$~ and $T_{A2}(x)$, the
first order phase transition to the AFE phase, with order
parameter $Q_{1}$, intersects the second order phase transition to
the AFE phase, with order parameter $Q_{2}$, at a \textit{critical
endpoint} [22]. The principle experimental evidence supporting
this interpretation is that the thermal hysteresis vanishes
abruptly at $x_{0}$ (Fig.4). Note that we assume $B_{21}>0$~ and
$B_{22}>0$~ but the sign ''-'' stands in (\ref{Free} ) in front of
$B_{21 }$~ and ''+'' precedes $B_{22}$. The alternative
assumption, that there is a tricritical point at ($x_0$,
$T_{Ai}(x)$), would imply a \textit{gradual disappearance} of the
thermal hysteresis as $x$~ approaches $x_{0}$~ from below.

Strong $x$-dependence of the thermal hysteresis at $x<x_{0}$~ is
unusual. From (\ref{eq8}) it follows that $B_{21}$~ increases with
$x$ regardless of the sign of the electrostriction constant. At
small $x$, this leads to a linear increase in $\Delta T_{h}(x)$~
while at larger $x$, $\Delta T_{h}(x)$~ increases quadratically,
in excellent agreement with experiment (Fig.4).

From (\ref{eq16}) it follows that the jump, $\Delta $Q$_{1}$, in
AFE order parameter $Q_{1 }$~ should depend on the concentration
$x$. Because $B_{2}$ increases linearly with $x$, so must $\Delta
Q_{1}$. Experimentally, it is observed that as $x$~ increases, so
does the difference between the magnitudes of $\varepsilon '_{m}$~
measured on cooling and on heating (Fig.1), which implies good
agreement between theory and experiment.

Our data are insufficient to identify the crystal structure of the
AFE phase associated with $Q_{2}$=\textit{finite }($x_{0}<x$) but
some properties are clear: $\varepsilon '(T)$~ is relaxor- like
with a diffuse maximum and dispersion below $T_{m}$. Some
properties however, are not typically relaxor- like: the
dielectric permittivity magnitudes are much lower than those in
typical relaxors such as PMN; the $\varepsilon $'$(T)$~ fits the
Curie-Weiss law well above $T_{m}$, but the extrapolated
Curie-Weiss temperature, $T_{CW}$, is significantly below $T_{m}$~
rather than above as in typical relaxors; in spite of the dramatic
diffusion of $\varepsilon '(T)$, the frequency dependence of
$\varepsilon '$~ and $T_{m}$~ is much weaker than in typical
relaxors. The comparatively small magnitude of $\varepsilon
'_{m}$~ can be connected with the absence of the large
contribution of lead to the dielectric permittivity.

\section{SUMMARY}

From the discussion above it follows that one of the main reasons
for the appearance of the relaxor-like properties in the
NaNbO$_{3}$-based solid solutions belonging to group II can be the
appearance of the (random) local strain (with nonvanishing average
magnitude) stemmed from the impurities. Due to the
electrostriction effect this results in the decrease of the AFE
critical temperature, increase of the thermal hysteresis width
with x, and, finally, with the appearance of a new diffuse phase
transition with weak dispersion of the dielectric permittivity.
Large diffusion of the $\varepsilon '(T)$~ maxima in some solid
solution compositions with $x > x_{0}$~ and the possibility of
shifting $T_{m}$~ to the room temperature range in conjunction
with relatively weak frequency dependence of both $T_{m}$~ and
$\varepsilon '_{m}$~ may be of certain interest for applications.
The same is true for solid solution compositions with $x < x_{0}$~
exhibiting giant $\varepsilon '(T)$~ thermal hysteresis values.

\section{ACKNOWLEDGMENTS}

This work was partially supported by Russian Foundation for Basic
Research (Grants {\#} 01-03-33119 and 01-02-16029). S.A.P.
appreciates discussions with M. Glinchuk and B. Burton.

\end{document}